\documentclass[conference]{IEEEtran}
\IEEEoverridecommandlockouts
\usepackage{cite}
\usepackage{url}
\usepackage{booktabs}
\usepackage{amsmath,amssymb,amsfonts}
\usepackage{algorithmic}
\usepackage{graphicx}
\usepackage{textcomp}
\usepackage{xcolor}
\def\BibTeX{{\rm B\kern-.05em{\sc i\kern-.025em b}\kern-.08em
    T\kern-.1667em\lower.7ex\hbox{E}\kern-.125emX}}
\begin{document}

\title{SuperCaustics: Real-time, open-source simulation of transparent objects for deep learning applications \\
}

\author{\IEEEauthorblockN{1\textsuperscript{st} Mehdi Mousavi}
\IEEEauthorblockA{\textit{Department of Computer Science} \\
\textit{Georgia State University}\\
Atlanta, United States \\
smousavi2@student.gsu.edu}
\and
\IEEEauthorblockN{2\textsuperscript{nd} Rolando Estrada}
\IEEEauthorblockA{\textit{Department of Computer Science} \\
\textit{Georgia State University}\\
Atlanta, United States \\
restrada1@gsu.edu}

\\

}



\maketitle

\begin{abstract}
Transparent objects are a very challenging problem in computer vision. They are hard to segment or classify due to their lack of precise boundaries, and there is limited data available for training deep neural networks. As such, current solutions for this problem employ rigid synthetic datasets, which lack flexibility and lead to severe performance degradation when deployed on real-world scenarios. In particular, these synthetic datasets omit features such as refraction, dispersion and caustics due to limitations in the rendering pipeline. To address this issue, we present SuperCaustics, a real-time, open-source simulation of transparent objects designed for deep learning applications. SuperCaustics features extensive modules for stochastic environment creation; uses hardware ray-tracing to support caustics, dispersion, and refraction; and enables generating massive datasets with multi-modal, pixel-perfect ground truth annotations. To validate our proposed system, we trained a deep neural network from scratch to segment transparent objects in difficult lighting scenarios. Our neural network achieved performance comparable to the state-of-the-art on a real-world dataset using only 10\% of the training data and in a fraction of the training time. Further experiments show that a model trained with SuperCaustics can segment different types of caustics, even in images with multiple overlapping transparent objects. To the best of our knowledge, this is the first such result for a model trained on synthetic data. Both our open-source code and experimental data are freely available online. 
\end{abstract}


\begin{IEEEkeywords}
Deep Learning, Computer Vision, Synthetic Data, Dataset, AI Simulation, Virtual Environment, Caustics
\end{IEEEkeywords}

\section{Introduction}
Detecting transparent objects is one of the most challenging problems in computer vision, because these objects do not form disjoint boundaries with their environment. In addition, different types of transparent objects have unique characteristics based on the materials they're made of, and they come in varied thicknesses and densities that affects their appearance. Light passing through a transparent object will bounce multiple times before it reaches our eyes (or a camera), causing effects such as refraction, specular reflections, caustics and dispersion; and the dynamic nature of light makes them appear in radically different ways when exposed to light from different angles.

\begin{figure}

 \center
  \includegraphics[width=\linewidth]{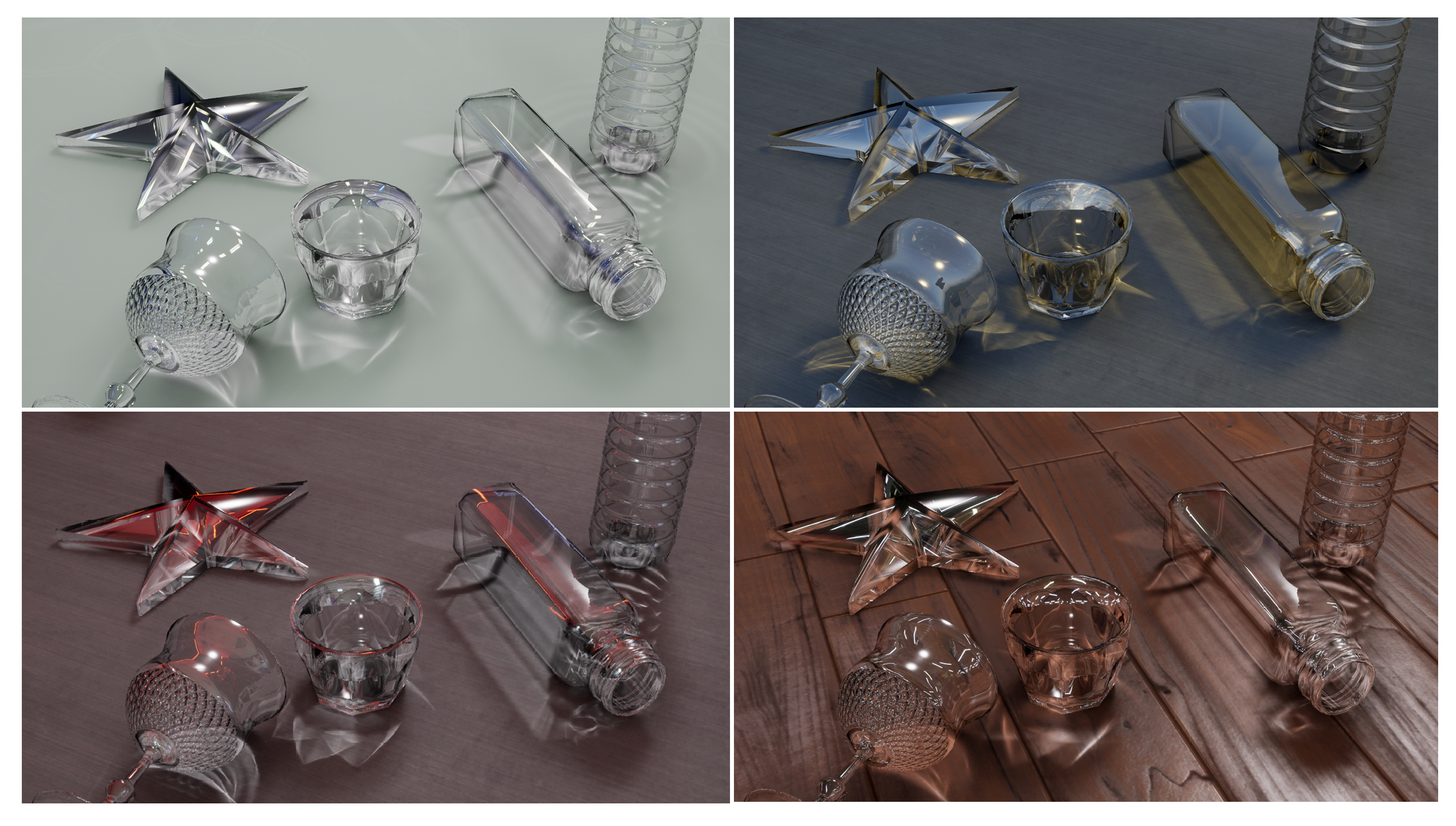}
  \caption{Images rendered in SuperCaustics. Changing image features in real time is trivial in SuperCaustics. Note the changes in lighting, sharpness of caustics, realistic backdrop and specular reflections on the transparent objects.} 

  \label{teas}

\end{figure}

Deep learning has flourished into a rapidly thriving field over the last decade. In particular, it has achieved state-of-the-art results in problems for which large amounts of (usually labelled) data is available. However, there has been relatively little work on applying deep learning to transparent object detection in part because gathering and labeling a sufficiently large dataset for this problem is cumbersome and difficult. Most recent publications on transparent object detection have utilized small, custom-made datasets of real data, gathered and labelled by the authors themselves to \textit{validate or test} their algorithm or DCNN model. The actual training data consists primarily of large volumes of synthetic images created in 3D modelling software (e.g., Blender \cite{blender}).


 

An important limitation of these rigid image datasets is their static nature. For instance, once an image has been generated, one cannot change its illumination, or replace the reflection maps in the environment. For real images, labeling the ground truth is also a significant challenge, especially for cameras with high pixel counts. \color{black} Depth, in particular, is almost impossible to label by hand, so one is limited to active depth sensing (e.g., RGB-D sensors) or passive methods like disparity matching. However, depth cameras often register errors when encountering transparent objects due to their reflective and refractive properties \cite{cleargrasp}. In computer vision applications where transparent objects are involved |such as robotic grasping| precise labeling is an integral part of the entire pipeline, and having fault-free data is crucial to the function of a robotics grasp algorithm. For example, the Cleargrasp pipeline \cite{cleargrasp}, the current state-of-the-art for deep-learning based transparent object detection, uses a segmentation map to complement and remove faulty depth signals from a depth image.




Moreover, no matter how extensive and large a pre-made training dataset is, it can only cover so many lighting conditions. The positions of objects, type of surfaces, camera angles, and placement of light sources will only cover a small subset of all possible combinations. This is a challenge for transparent objects because their appearance is highly dependent on how they are lit. For example, they might cast shadows, caustics, or a combination of the two simply by varying the angle or intensity of the light sources. As such, models trained on synthetic datasets with only a small range of lightning scenarios suffer performance degradation when deployed on real-world data. 

 

 

More specifically, as we review in Section~\ref{sec:related}, clever deep learning solutions are often stifled by rigid datasets that cannot represent the full possibility space of the target domain, are difficult to change after acquisition, require significant human effort to produce or use proprietary technology for which the source-code and project files used to generate them are rarely released for free use by the public.

To address this gap, we propose SuperCaustics, an easy to use, highly customizable, easily extendable, open-source real-time simulation tool designed for generating dynamic, massive datasets in highly complex scenarios. As we detail further in Section III, SuperCaustics is a modular set of customizable classes developed within NVIDIA's RTX Branch of Unreal Engine 4 (Epic Games, USA)\cite{unrealengine,nvidia-github-repository}, allowing even those researchers who are unfamiliar with Unreal Engine to create, maintain and iterate on their own photorealistic customized datasets for vision tasks involving transparent objects. SuperCaustics features four modules: 

\begin{enumerate}
    \item A stochastic scene generation system that creates the geometry of a scene based on parameters like object type, number of objects, camera type, lighting, and background.
    \item A prop manager that adds or removes customizable background items to increase scene complexity and variety.
    \item A ground truth core that automatically overlays a wide variety of accurate, pixelwise ground-truth annotations.
    \item A data ablation core \cite{AIP} that enables changing characteristics of the objects, background, environment, lighting, or rendering without affecting scene composition.
\end{enumerate}

\begin{figure}

 \center

  \includegraphics[width=\linewidth]{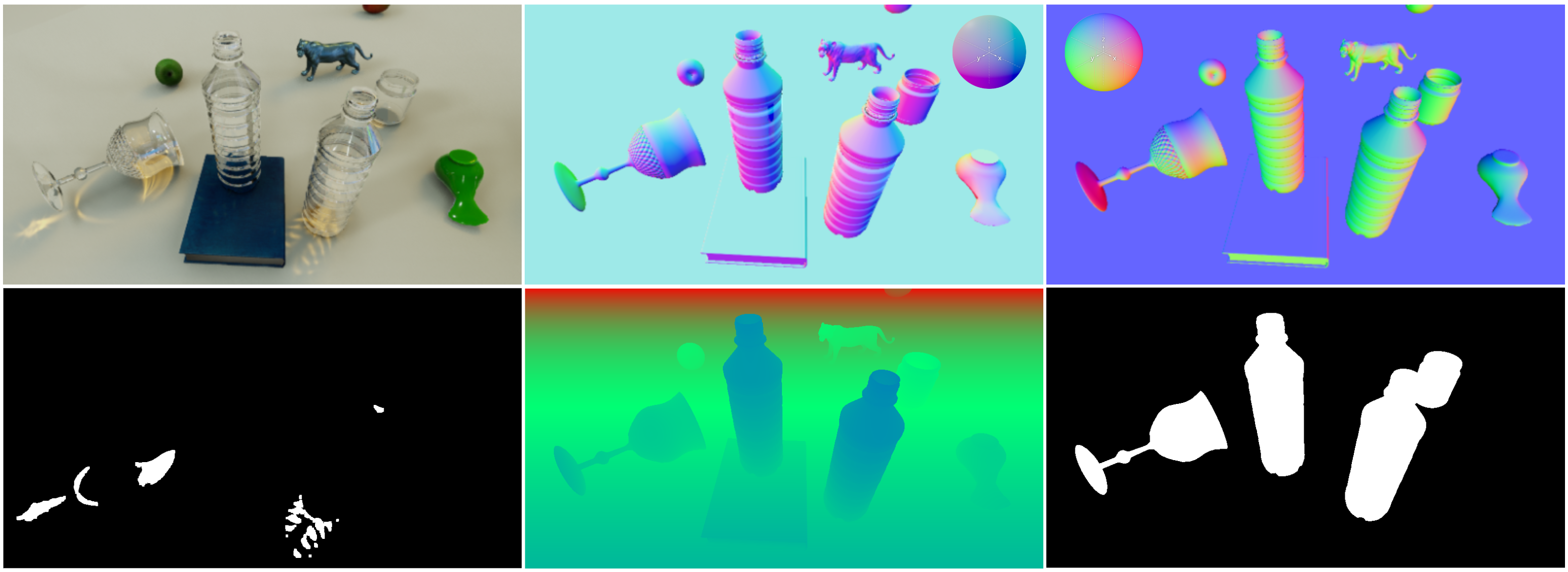}

  \caption{Showcase of some of the available ground truth images in SuperCaustics. First row (from left to right): caustics-enabled image, camera-space normals, world-space normals.
  Second row (from left to right): caustics segmentation, depth image, segmentation mask. (Image best viewed on screen.)} 

  \label{GTShow}

\end{figure}

\begin{figure*}

 \center

  \includegraphics[width=\textwidth]{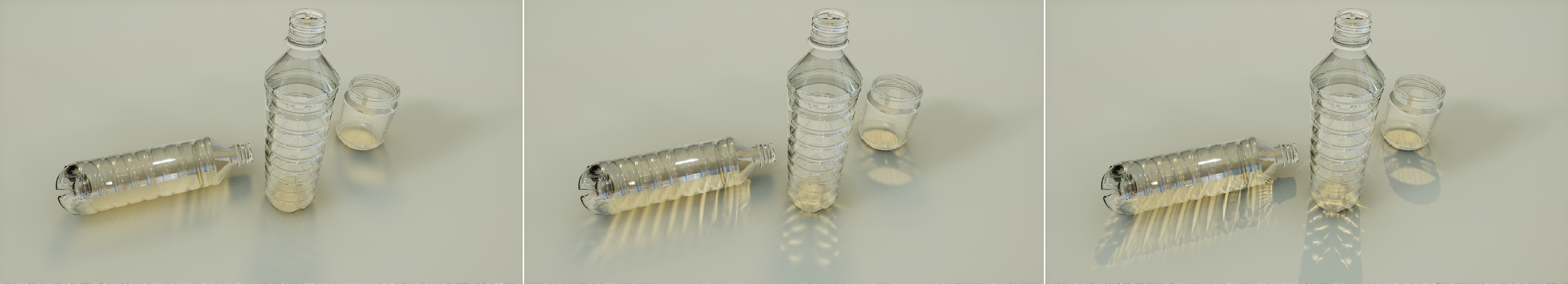}

  \caption{SuperCaustics allows fine controls over fundamental characteristics of every variable in the simulation. e.g., controlling the softness of the light-source and its caustics. Left: Very soft caustics, Middle: Moderately soft, Right: Very Sharp.}

  \label{Softness}

\end{figure*}

Our proposed system allows for fine granular control over small details of the scene before and after generation. Each module has a set of customizable and extendable controls, including tools for data ablation. As detailed in \cite{AIP}, data ablation is the study of the effects of isolated changes in the features of data on the performance of a deep learning network (e.g. having the exact same image under a different lighting condition). Using the Data Ablation Core, we are able to change the characteristics of a scene in real-time, e.g., different lighting color, presence of sharp or soft caustics, light source angle, texture or shape of the backdrop, or changes in roughness/color of the glass surfaces (see Fig.~\ref{Softness}). The tools in the Data Ablation Core can also be used to increase variety in the captured images. 

A virtual dataset generated from SuperCaustics contains the features of the target domain, free from acquisition errors or labeling bias. More importantly, as our experiments confirm, a small, targeted, carefully crafted data-set can efficiently match the performance achieved by state-of-the-art methods that train their systems on massive datasets.

\begin{figure*}

 \center

  \includegraphics[width=.75\textwidth]{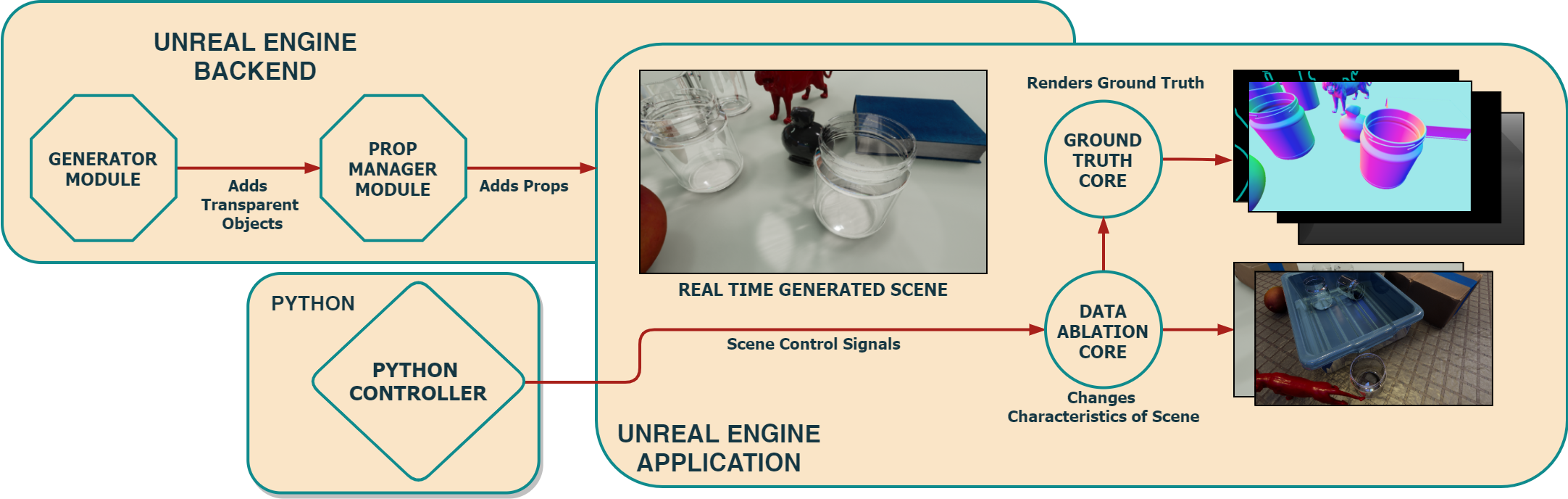}

  \caption{As illustrated, our system has four components: (1) Scene Generator Module; (2) Prop Manager Module; (3) Ground-truth Core; (4) Data Ablation Core. Each of these modules comes with its own sub-modules that enable generation of a customizable scene in real-time. We use Blueprint, Unreal's visual scripting language, for the ground-truth annotations and data ablation controls. We use a separate Python interaction module|Controller|for data collection. }

  \label{Diagram}

\end{figure*}

\section{Related Work}\label{sec:related}
\noindent\textbf{The old and the new - A problem that persists: }Recognizing transparent objects has been a difficult problem to solve in vision. Early solutions used general features of transparent objects, e.g., specular reflections, for object recognition \cite{specular}. More sophisticated approaches include estimating bounding boxes based on the refraction distortions that transparent objects' edges create on the background \cite{additive}. Newer research trains deep convolutional neural networks on massive datasets to reach a pixel-wise labeling of transparent objects \cite{cleargrasp, LocalImplicit}. Also, some researchers have devised clever methods to exploit sensor failures in depth images to localize transparent objects \cite{friendorfoe}. Regardless, often times researchers are forced to use low-quality real data for testing their algorithm or training a neural network \cite{cleargrasp,additive}, and other times researchers resort to creating rigid virtual datasets to train large models, such as a DCNN. \cite{LocalImplicit,friendorfoe,cleargrasp}.\\

\noindent \textbf{Synthetic Data: } Synthetic data has proven useful in various computer vision tasks, such as depth estimation \cite{soccer,cleargrasp}, surface normal estimation \cite{AIP,cleargrasp}, robotic grasping \cite{dexnet}, and object segmentation \cite{model-driven-sim} by multiple independent researchers \cite{Photorealism2,model-driven-sim,photorealism3,AIP}. Modern computer graphics can achieve near-photorealism, so synthetic data has become a viable alternative in situations where acquiring or labeling real data is difficult. However, there are still few datasets that meet the image quality required to be considered photo-realistic \cite{LocalImplicit}, and those that do, rarely include transparent objects because of the complexities they introduce in rendering and generating correct ground-truth labels. More importantly, to the best of our knowledge, there is no photorealistic, synthetic dataset that includes the tools and project files needed to generate the data itself. More often, these synthetic datasets are static and use proprietary code and technologies that does not allow for reproduction or change in the dataset. We believe flexibility in creating the image data is crucial since it allows for greater degrees of freedom when evaluating a specific method, and allows researchers to make corrections in the data and address weak spots in the model's performance.\\


\noindent \textbf{The need for reliable data:} One common factor in transparent object detection is the lack of reliable data. In older research, we see the use of personally-gathered real data that suffers from lack of variety, noise, or lack of labels. To get past this obstacle, we see various works generate their own synthetic datasets using tools like Blender \cite{cleargrasp,blender} or NVIDIA Omniverse \cite{LocalImplicit,OmniVerse}. As a result, these datasets consist of huge collections of static images. In addition, the developers of these datasets rarely release the project files or source code used to generate the data. Nevertheless, in most cases the target domain of the final algorithm is the real world. As such, to be able to generalize well into the target domain these datasets have to have tens of thousands of images, at least, making the training very expensive and time consuming. Despite all this data, we still see significant performance degradation when shifting from crisp, high-resolution synthetic data to blurry, grainy real images that are lit entirely differently. These synthetic datasets are non-changeable (static) and more often than not, they do not include graphically intense phenomena that happen in real-world scenarios (e.g., dispersion or caustics), which makes them very narrow in domain and makes the trained algorithms sensitive to these lighting conditions. \cite{LocalImplicit, cleargrasp}.\\

\noindent \textbf{SuperCaustics vs. Cleargrasp:} Arguably, the most significant effort in modeling transparent objects is Cleargrasp \cite{cleargrasp}. In this study, researchers trained deep learning models to estimate depth, surface normals, and occlusion boundaries using a mix of synthetic and real data. Their results indicate that in addition to a significant performance loss when shifting from a synthetic to a real domain, the models consistently misclassify caustics. In particular, they wrongly classify caustics as separate transparent objects. The authors of Cleargrasp speculated that this was because the same amount of sharp caustics were not present in the synthetic dataset due to limitations in the rendering software. A more flexible simulation tool, such as our proposed system, allows for addition or removal of such features. In particular, SuperCaustics can produce images with sharp caustics. More generally, our system can incorporate features of the real domain in the training data to make trained models more robust to such features in the target domain.\\

Overall, prior work on transparent object detection has been hampered by a lack of reliable, flexible data. To address this gap, we have developed SuperCaustics, an extendable, open-source, user-friendly simulation tool that allows for fine controls over every variable in the simulation. Our system can generate millions of unique images out of the box. In addition, it allows researchers to import 3D models of their choice, set their desired parameters, and capture synthetic data from fully customizable virtual environments. We describe our system in more detail, below.

\section{SuperCaustics}
\label{sec:SuperCaustics}
SuperCaustics is a simulation tool made in Unreal Engine for generating massive computer vision datasets that include transparent objects. Unreal Engine is the engine of choice for projects with high-resolution, real-time 3D graphics. It is free for both commercial and non-commercial use and its source code is publicly available and extended by the community. We use one such extended version created for hardware raytracing by NVIDIA RTX graphics cards \cite{nvidia-github-repository}.

The key contributions of SuperCaustics are: \textbf{1.} Open-source, free set of user-friendly, extendable tools for creating custom datasets.
\textbf{2.} Customizable photorealistic scenes using real-time hardware ray-tracing,
\textbf{3.} Automatic, multimodal, accurate ground-truth generation for visual tasks,
\textbf{4.} Python and UE4 interplay for data acquisition and processing.

Below, we detail each major module of SuperCaustics, and provide explanations of the different ground-truth annotations available in SuperCaustics.

\subsection{Generator Module}\label{Generator}
The Generator Module creates a scene based on given parameters like item type, shape, number of items and/or range of items. It leverages the physics engine inside UE4 to drop the items on the scene within a customizable distance offset. A random physics impulse can be added to the objects upon spawning to increase randomness and add variety to how they come to rest. The intensity of the impulse is also customizable.


\subsection{Props Manager}\label{Props}
Similar to the Generator Module, a separate module manages the background items in the simulation. This module accepts different 3D objects as prop inputs, and manages their position and rotation within the first few frames of the simulation until they come to rest by the physics engine. 

\subsection{Data Ablation Core}\label{DataAblation} Data Ablation is a powerful method for determining weak points and improving robustness in a neural network's performance \cite{AIP}. The Data Ablation Core in SuperCaustics allows for such experiments within the simulation. Additionally, the features of the Data Ablation Core can be used to further increase the variety in the final dataset. We can adjust the cameras (camera matrix), lighting, backdrops, reflection profiles (HDRI maps), and properties of the glass material (color, specular, thickness, roughness, texture). Additionally, all properties of the ray-tracing engine are exposed for experimentation within the Data Ablation module.

\subsection{Ground Truth Core}\label{GT}

SuperCaustics also includes scripts for automatically estimating depth, surface normals (world space and camera space), object masks, outlines, occlusion boundaries and caustics segmentation. The Ground Truth Core can be readily extended by adding additional scripts (or modifying the available scripts). We use post-processing shaders (dubbed \textit{materials} in UE4), to overlay these properties over the image, enabling pixel-perfect alignment between the data and the ground-truth labels. (See Figure \ref{GTShow}) We detail each ground-truth mode below:

\subsubsection{\textbf{Depth Ground Truth}} We calculate the normalized distance between each pixel that belongs to a specific object and the camera. By default, we set the real-life range of depth to 10 meters, which covers the entire environment. This range is customizable in the Ground Truth Core.

\subsubsection{\textbf{Surface Normals Ground Truth}} We estimate the normal vector w.r.t to each 3D surface, then color each pixel to indicate the vector's direction. We use 6 main colors to show 6 axis of direction (positive and negative xyz). Surface normals can be calculated in 2 different spaces: camera-space and world-space. Normals of a pixel in world-space are defined with respect to the world Cartesian coordinate system. In world-space, each normal vector points towards a fixed axis in space, regardless of where it is being observed from. Camera-space is the space in which points are defined with respect to the camera coordinate system. Each normal vector has the transformation of the camera itself applied to it, making the camera the origin point of the coordinate system. This means the transformations of the camera in world-space is implicitly applied to every normal vector in the image. This is a simplification compared to world-space surface normals. (See Figure \ref{GTShow}.)

\begin{table*}[!h]
\centering
\caption{Models trained with SuperCaustics Data generalize well to novel scenarios}
\label{tab:cross}
\resizebox{\textwidth}{!}{%
\begin{tabular}{lllcccc}
\hline
\multicolumn{2}{c}{\textbf{Experiment}}         & \multicolumn{5}{c}{\textbf{Scores Obtained}} \\ \hline
\textbf{Training} & \textbf{Testing} & \textbf{Accuracy} & \textbf{F1 Score} & \textbf{Precision} & \textbf{Recall} & \textbf{IOU} \\
SuperCaustics   & SuperCaustics (Known Objects) & 0.9924  & 0.9361  & 0.9975 & 0.8818 & 0.8801 \\
SuperCaustics-R & Cleargrasp-Real (Novel)       & 0.9560  & 0.6941  & 0.5627 & 0.9056 & 0.5316 \\
ClearGrasp      & Cleargrasp-Real (Novel)       & NR      & NR      & NR     & NR     & 0.5800 \\ \hline
\end{tabular}%
}
\end{table*}

\subsubsection{\textbf{Object Mask Ground Truth}} In Unreal Engine, it is easy to map visible pixels to their corresponding 3D objects. Our Blueprint script uses this mapping to automatically detect and overlay pixel-perfect semantic labels on the objects in the scene.

\subsubsection{\textbf{Outlines Ground Truth}} The edges of transparent objects are sometimes hard to detect visually. In this Ground Truth pass, we can highlight the edges of objects with a customizable thickness (e.g., 1px or 20px).

\subsubsection{\textbf{Occlusion Boundaries Ground Truth}} This Ground Truth pass shows surfaces where transparent objects are touching another surface (like the backdrop). Specifically, this pass shows where there is a depth discontinuity. The logic of this ground truth pass is adapted from Cleargrasp \cite{cleargrasp}.

\subsubsection{\textbf {Local \& Non-local Caustics Mask Ground Truth}} In SuperCaustics, we have fine control over the characteristics of the image, and we  can create the exact same image under different settings. We use the difference in the  luminosity between two rendered images (one with caustics, and another without) to reach a pixel-wise segmentation of local caustics (inside transparent object) and non-local caustics (projected unto another surface) in the image.


\section{Evaluation}
To validate the usefulness of SuperCaustics, we trained a deep convolutional neural network (DCNN) to do a pixel-wise labeling of transparent objects with both soft and sharp caustics. We chose pixel-wise labeling because it was more suitable given the limited hardware we had available. To demonstrate the flexibility of SuperCaustics, we aimed to \textit{test} our model on the real test set released by Cleargrasp. To do that, we set the angles of the simulated Intel real-sense cameras to resemble similar perspectives and based the parameters of our simulation on the characteristics of the Cleargrasp's real-world dataset (i.e., occasional sharp caustics from hard lights, general soft lighting, presence of a tote box and wooden and cloth backdrops among synthetic backgrounds). Below, we compare our network's performance to the network proposed in Cleargrasp \cite{cleargrasp}. Our results indicate that our network achieved comparable results with much fewer training data.


\subsection{Experimental Setup}

\subsubsection{Hardware} We conducted all our experiments in a Dell Precision 7920R server with  two  Intel  Xeon  Silver  4110  CPUs,  two  GeForce  GTX  1080  Ti  graphics cards, and 128 GBs of RAM.

\subsubsection{Image acquisition} Using the features described in Section~\ref{sec:SuperCaustics}, we generated 9000 1920$\times$1080 synthetic images from 650 randomized scenarios. In each scenario, a reflection map and a backdrop was randomly chosen from a bank of 33 HDRI mappings and 33 backdrops. Then, a number of transparent objects were dropped into the scene from off-camera, so that they came to rest naturally using UE4's physics engine. Then, the Prop Manager module added the input props in random locations inside the scene. For gathering the images, 12 Intel Realsense cameras were simulated in various locations and angles. In every camera angle, the Data Ablation Core rotated the main light of the scene to cast shadows and caustics at various angles before capturing an image and its ground-truth values. For training, the images were split in the following ratio: 8000 for training and 1000 for validation.

\subsubsection{Deep Neural Networks} We used an implementation of U-net \cite{unet} as our segmentation neural network. In particular, we used the tools provided by the EasyTorch Library to prototype and run the experiments. EasyTorch is a Pytorch-based deep learning library used in \cite{easytorch,10.3389/fcomp.2020.00035,motevali}. We evaluated our performance on mean intersection-over-union (IOU) for the test set. For the loss function, we used the negative log likelihood loss (NLL). We trained our network for 30 epochs on purely synthetic data from SuperCaustics. Afterwards, we used the SuperCaustics model as a pre-trained starting point (labeled SuperCaustics-R in our results), which was trained on Cleargrasp Known Objects for an additional four epochs. 

\begin{table*}[]
\centering
\caption{Caustics Segmentation}
\label{tab:caustics}
\resizebox{\textwidth}{!}{%
\begin{tabular}{@{}llccccc@{}}
\toprule
\multicolumn{2}{c}{\textbf{Experiment}}                     & \multicolumn{5}{c}{\textbf{Scores Obtained}} \\ \midrule
\textbf{Training} & \textbf{Testing} & \textbf{Accuracy} & \textbf{F1 Score} & \textbf{Precision} & \textbf{Recall} & \textbf{IOU} \\
Caustic Segmentation & Caustic Segmentation (Known Objects) & 0.9981  & 0.9386  & 0.9541 & 0.9237 & 0.8844 \\
Caustic Segmentation & Caustic Segmentation (Novel Objects) & 0.9958  & 0.8102  & 0.8682 & 0.7594 & 0.6809 \\ \bottomrule
\end{tabular}%
}
\end{table*}

\subsection{Results \& Discussion}
Table~\ref{tab:cross} shows the object segmentation performance of our network trained on SuperCaustics data, and Table~\ref{tab:caustics} shows the equivalent results when segmenting caustics directly. In particular, the first table shows that our SuperCaustics-R model (which was also trained on a subset of real data from Cleargrasp) generalizes well to novel objects. Our network trained only with synthetic data achieves 88\% IOU on the same domain, and it is robust to challenging image features like sharp caustics and segmenting multiple overlapping objects. Moreover, our SuperCaustics-R model achieves a performance of 53\% IOU (comparable to Cleargrasp's 58\%) on the Novel Objects test set. Below, we analyze these results in more detail.


  \begin{figure}
    \center
    \includegraphics[width=\linewidth]{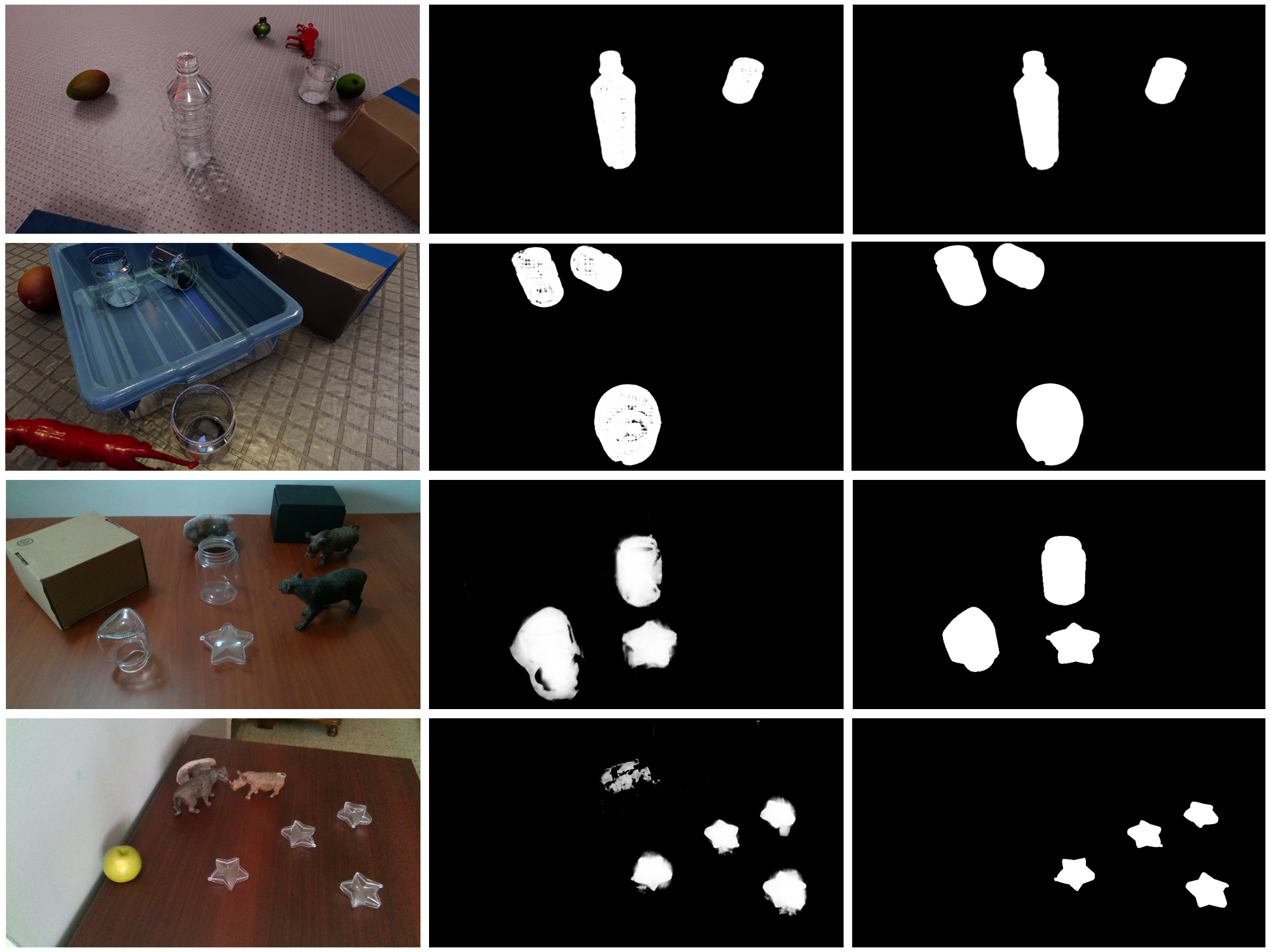}
    \caption{Image (left), Prediction (middle), Ground-truth (right) First two rows show SuperCaustics Model on SuperCaustics data. Second two rows show SuperCaustics-R on Cleargrasp-Novel.}\label{show-novel}
  \end{figure}
  
\noindent{\textbf{Performance, Efficiency and Scalability:}} Compared to \cite{cleargrasp}, our training procedure is much more efficient. By producing higher quality, wider-domain data, we can achieve roughly the same performance on the Novel Objects test set using only 10\% of the amount of data and in a fraction of the training time used by Cleargrasp (34 epochs on about 8000 images, compared to 100 epochs on 80000 images). Moreover, our training setup did not require 8 Tesla v100 cards as used by Cleargrasp. Achieving comparable results more efficiently supports our assertion that for difficult, complex tasks like labeling shiny transparent objects with caustics, using a rigid, massive dataset is not ideal, and one can always benefit from the availability of a better curated, higher quality dataset. That is why open-source, user-friendly simulations like SuperCaustics are important, because for every use-case scenario, a well-made custom dataset is going to achieve comparable results quicker and cheaper than traditional, massive datasets.\\


\noindent{\textbf{Robustness to Caustics: }} Fig.~\ref{show-novel} shows examples of images with sharp caustics that were produced by our SuperCaustics system. Our trained model does not confuse caustics with separate transparent objects and is quite robust to caustics from various angles and light sources. This is due to the presence of natural caustics in the dataset.\\

\noindent{\textbf{Performance on the Real Domain: }} As Fig.~\ref{show-novel} shows, our trained model seems to have learned the general characteristics of transparent objects from the SuperCaustics synthetic data. However, as opposed to Cleagrasp where known objects were present in the synthetic data, our model was not exposed to the shape of these novel objects. Performance in the real domain was also affected by the grainy, noisy nature of the real-camera. Prior work has shown that segmentation models are very sensitive to fidelity \cite{AIP}, thus, a change in the general fidelity of the domain had a significant, but expected impact on the performance of the model.\\

\noindent{\textbf{A Solution to Cleargrasp's main problem | Segmenting Non-local Caustics: }} In Cleargrasp's original experiments, their trained models falsely classify caustics as separate transparent objects \cite{cleargrasp}. One possible solution for this problem is having a separate system that only classifies caustics, and because predictions in these regions tend to be invalid, masking the caustics out of the predictions. This setup is straightforward with SuperCaustics. To test this, we generated 2300 1280$\times$720 images using the caustics segmentation mask ground-truth. We then trained a segmentation model to identify sharp, refractive, and reflective non-local caustics for 50 epochs. We then generated 60 images using novel objects (not seen before by the model) to test the generalization of the model.  As Table~\ref{tab:caustics} shows, our model was able to segment caustics cast by novel objects at 68\% IOU rate. As such, we believe that training a model with our SuperCaustics system will prove a feasible solution for masking out the erroneous pixels in Cleargrasp.\\ \\

  \begin{figure}
    \center
    \includegraphics[width=.8\linewidth]{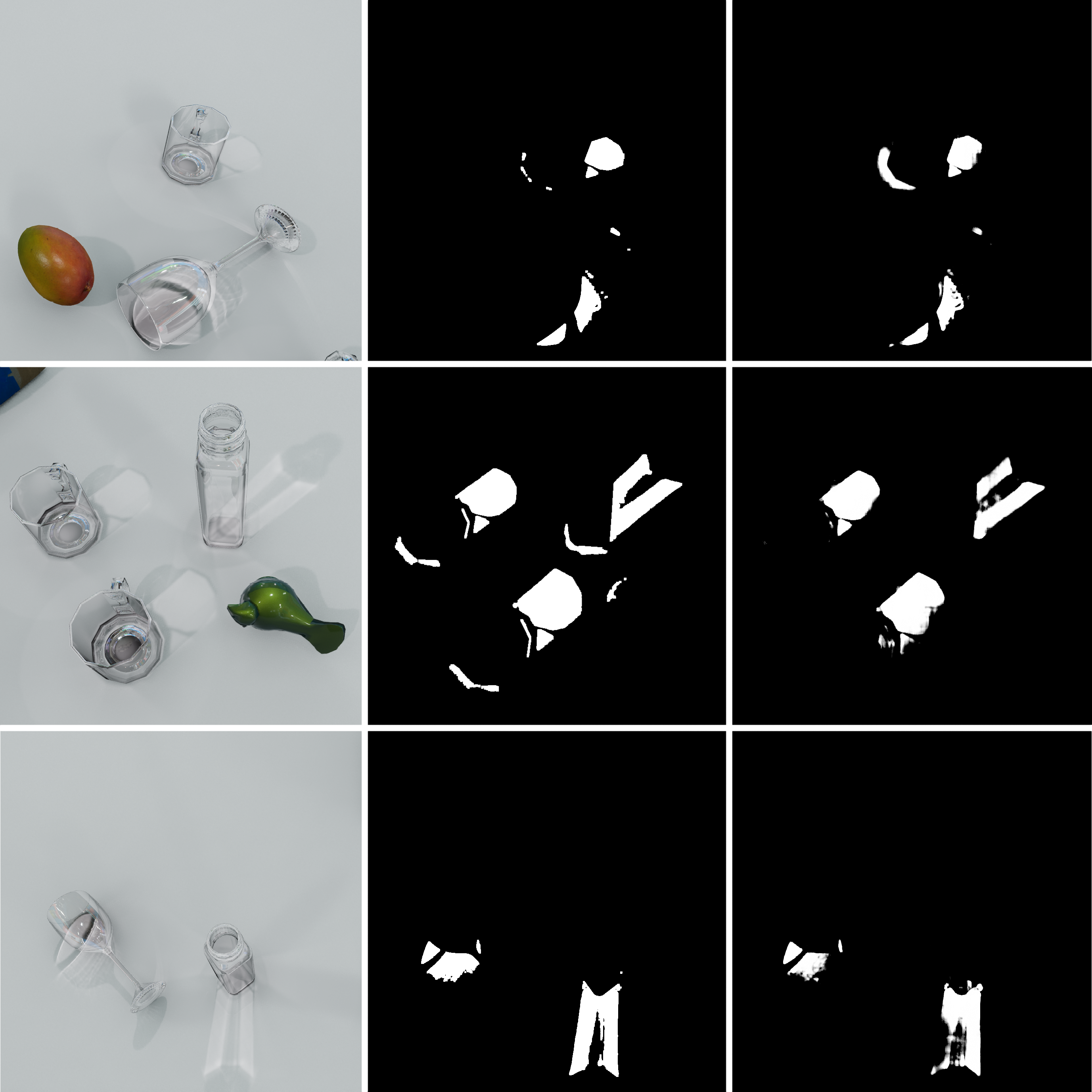}
    \caption{Image (left), Ground-truth (middle), Prediction (right),  SuperCaustics Model on Segmenting Caustics cast by Novel transparent objects.}\label{caustic-seg}
  \end{figure}

\noindent{\textbf{Sensitivity to Brightness:}} Our trained model seems to be very sensitive to brightness in the features of the objects. In our tests with synthetic data, where the camera's exposure underexposes shadows, our model seems to make mistakes in segmenting transparent objects.\\

\noindent{\textbf{Side-Effect of the Patch Based Approach:}} In some of the images, there is some noise visible in the labeled pixels. We believe this is due to the patch-based approach used by the U-Net neural network; specifically, the noise is a result of the artifacts from stitching these patches together in the final image. In most cases, patches are surrounded by uniform neighboring pixels, which could be used as information in a context-based post processing algorithm to get rid of the noise.

\section{Conclusion \& Future Work}

\noindent{\textbf{Conclusion: }}In this paper, we present SuperCaustics, an extendable, open-source, and user-friendly simulation tool for transparent objects that allows for fine controls over every variable in the simulation. Our system allows users to create customized datasets using its flexible simulation tools. With SuperCaustics, we hope to address the dearth of high-quality data for transparent objects, particularly data that can be modified post-acquisition. We demonstrated that by using a better curated data, a network can achieve performance comparable to when training with much larger rigid datasets. In particular, we were able to achieve comparable performance to the state-of-the-art networks trained on Cleargrasp's data using a dataset of only 9000 RGB images. Specifically, we found that high-quality data is linked to better performance in segmentation, as our SuperCaustics-R model was able to approach the performance of state-of-the-art (cleargrasp) with only 34 epochs of training and 10\% of the data, on far more reasonable hardware.

We believe that SuperCaustics will open exciting avenues for using synthetic data in machine learning, particularly for vision research related to transparent objects.\\

\noindent{\textbf{Future Work: }}
Potential research possibilities enabled by the availability of readily-expandable, open-source simulations are endless. However, we are working on improving and developing our own deep learning pipeline for other tasks such as surface normal estimation and occlusion-boundary detection. As we have shown, an exciting avenue that can make full use of SuperCaustics is classifying projected local and non-local caustics in images of transparent objects. This can be useful in complex pipelines that are sensitive to caustics and need a masking pass to make sure only transparent objects are processed when estimating surface normals, depth, or segmentation boundaries. Other avenues for future research are performing data ablation studies, where isolated features in the data are changed to determine sensitive points in the performance of a specific neural network, and using the class based features developed in SuperCaustics to help solve problems such as active learning, lifelong learning and class incremental learning \cite{10.3389/frai.2020.00019,DJay,blake1,jay3}. Finally, we will continue developing tools to expand the capabilities of our SuperCaustics system, as well as improving its stability and compatibility with different target domains. 


\bibliographystyle{./bibliography/IEEEtran}
\bibliography{./bibliography/IEEEabrv,./bibliography/IEEEexample}

\begin{thebibliography}{10}
\providecommand{\url}[1]{#1}
\csname url@samestyle\endcsname
\providecommand{\newblock}{\relax}
\providecommand{\bibinfo}[2]{#2}
\providecommand{\BIBentrySTDinterwordspacing}{\spaceskip=0pt\relax}
\providecommand{\BIBentryALTinterwordstretchfactor}{4}
\providecommand{\BIBentryALTinterwordspacing}{\spaceskip=\fontdimen2\font plus
\BIBentryALTinterwordstretchfactor\fontdimen3\font minus
  \fontdimen4\font\relax}
\providecommand{\BIBforeignlanguage}[2]{{%
\expandafter\ifx\csname l@#1\endcsname\relax
\typeout{** WARNING: IEEEtran.bst: No hyphenation pattern has been}%
\typeout{** loaded for the language `#1'. Using the pattern for}%
\typeout{** the default language instead.}%
\else
\language=\csname l@#1\endcsname
\fi
#2}}
\providecommand{\BIBdecl}{\relax}
\BIBdecl

\bibitem{blender}
\BIBentryALTinterwordspacing
B.~O. Community, \emph{Blender - a 3D modelling and rendering package}, Blender
  Foundation, Stichting Blender Foundation, Amsterdam, 2018. [Online].
  Available: \url{http://www.blender.org}
\BIBentrySTDinterwordspacing

\bibitem{cleargrasp}
S.~S. Sajjan, M.~Moore, M.~Pan, G.~Nagaraja, J.~Lee, A.~Zeng, and S.~Song,
  ``Cleargrasp: 3d shape estimation of transparent objects for manipulation,''
  2019.

\bibitem{unrealengine}
\BIBentryALTinterwordspacing
{Epic Games}, ``Unreal engine.'' [Online]. Available:
  \url{https://www.unrealengine.com}
\BIBentrySTDinterwordspacing

\bibitem{nvidia-github-repository}
\BIBentryALTinterwordspacing
NVIDIA, ``Nv{RTX} branch of {U}nreal {E}ngine,'' 2021. [Online]. Available:
  \url{https://github.com/NvRTX/UnrealEngine/tree/NvRTX\_Caustics-4.26}
\BIBentrySTDinterwordspacing

\bibitem{AIP}
M.~Mousavi, A.~Khanal, and R.~Estrada, ``Ai playground: Unreal engine-based
  data ablation tool for deep learning,'' in \emph{International Symposium on
  Visual Computing}.\hskip 1em plus 0.5em minus 0.4em\relax Springer, 2020, pp.
  518--532.

\bibitem{specular}
Osadchy, Jacobs, and Ramamoorthi, ``Using specularities for recognition,'' in
  \emph{Proceedings Ninth IEEE International Conference on Computer Vision},
  2003, pp. 1512--1519 vol.2.

\bibitem{additive}
M.~Fritz, G.~Bradski, S.~Karayev, T.~Darrell, and M.~Black, ``An additive
  latent feature model for transparent object recognition,'' \emph{Advances in
  Neural Information Processing Systems}, vol.~22, pp. 558--566, 2009.

\bibitem{LocalImplicit}
\BIBentryALTinterwordspacing
L.~Zhu, A.~Mousavian, Y.~Xiang, H.~Mazhar, J.~van Eenbergen, S.~Debnath, and
  D.~Fox, ``{RGB-D} local implicit function for depth completion of transparent
  objects,'' \emph{CoRR}, vol. abs/2104.00622, 2021. [Online]. Available:
  \url{https://arxiv.org/abs/2104.00622}
\BIBentrySTDinterwordspacing

\bibitem{friendorfoe}
\BIBentryALTinterwordspacing
V.~Seib, A.~Barthen, P.~Marohn, and D.~Paulus, ``{Friend or foe: exploiting
  sensor failures for transparent object localization and classification},'' in
  \emph{2016 International Conference on Robotics and Machine Vision}, A.~V.
  Bernstein, A.~Olaru, and J.~Zhou, Eds., vol. 10253, International Society for
  Optics and Photonics.\hskip 1em plus 0.5em minus 0.4em\relax SPIE, 2017, pp.
  94 -- 98. [Online]. Available: \url{https://doi.org/10.1117/12.2266255}
\BIBentrySTDinterwordspacing

\bibitem{soccer}
\BIBentryALTinterwordspacing
K.~Rematas, I.~Kemelmacher{-}Shlizerman, B.~Curless, and S.~M. Seitz, ``Soccer
  on your tabletop,'' \emph{CoRR}, vol. abs/1806.00890, 2018. [Online].
  Available: \url{http://arxiv.org/abs/1806.00890}
\BIBentrySTDinterwordspacing

\bibitem{dexnet}
J.~Mahler, J.~Liang, S.~Niyaz, M.~Laskey, R.~Doan, X.~Liu, J.~A. Ojea, and
  K.~Goldberg, ``Dex-net 2.0: Deep learning to plan robust grasps with
  synthetic point clouds and analytic grasp metrics,'' 2017.

\bibitem{model-driven-sim}
V.~{Veeravasarapu}, C.~{Rothkopf}, and R.~{Visvanathan}, ``Model-driven
  simulations for computer vision,'' in \emph{2017 IEEE Winter Conference on
  Applications of Computer Vision (WACV)}, 2017, pp. 1063--1071.

\bibitem{Photorealism2}
V.~Haltakov, C.~Unger, and S.~Ilic, ``Framework for generation of synthetic
  ground truth data for driver assistance applications,'' in \emph{GCPR}, 2013.

\bibitem{photorealism3}
A.~Gaidon, Q.~Wang, Y.~Cabon, and E.~Vig, ``Virtual worlds as proxy for
  multi-object tracking analysis,'' 2016.

\bibitem{OmniVerse}
\BIBentryALTinterwordspacing
{NVIDIA}. Nvidia omniverse platform. [Online]. Available:
  \url{https://developer.nvidia.com/nvidia-omniverse-platform}
\BIBentrySTDinterwordspacing

\bibitem{unet}
O.~Ronneberger, P.~Fischer, and T.~Brox, ``U-net: Convolutional networks for
  biomedical image segmentation,'' in \emph{International Conference on Medical
  image computing and computer-assisted intervention}.\hskip 1em plus 0.5em
  minus 0.4em\relax Springer, 2015, pp. 234--241.

\bibitem{easytorch}
A.~Khanal, ``Easy torch,'' \url{https://github.com/sraashis/easytorch}, 2020.

\bibitem{10.3389/fcomp.2020.00035}
\BIBentryALTinterwordspacing
A.~Khanal and R.~Estrada, ``Dynamic deep networks for retinal vessel
  segmentation,'' \emph{Frontiers in Computer Science}, vol.~2, p.~35, 2020.
  [Online]. Available:
  \url{https://www.frontiersin.org/article/10.3389/fcomp.2020.00035%7D,
  DOI={10.3389/fcomp.2020.00035}, ISSN={2624-9898}}
\BIBentrySTDinterwordspacing

\bibitem{motevali}
S.~Motevali, A.~Khanal, and R.~Estrada, ``Optic disc segmentation using
  disk-centered patch augmentation,'' 2021.

\bibitem{10.3389/frai.2020.00019}
\BIBentryALTinterwordspacing
J.~K. Mandivarapu, B.~Camp, and R.~Estrada, ``Self-net: Lifelong learning via
  continual self-modeling,'' \emph{Frontiers in Artificial Intelligence},
  vol.~3, p.~19, 2020. [Online]. Available:
  \url{https://www.frontiersin.org/article/10.3389/frai.2020.00019}
\BIBentrySTDinterwordspacing

\bibitem{DJay}
\BIBentryALTinterwordspacing
J.~Mandivarapu, B.~Camp, and R.~Estrada, ``Deep active learning via open set
  recognition,'' \emph{CoRR}, vol. abs/2007.02196, 2020. [Online]. Available:
  \url{https://arxiv.org/abs/2007.02196}
\BIBentrySTDinterwordspacing

\bibitem{blake1}
B.~Camp, J.~K. Mandivarapu, and R.~Estrada, ``Continual learning with deep
  artificial neurons,'' 2020.

\bibitem{jay3}
J.~K. Mandivarapu, E.~Bunch, Q.~You, and G.~Fung, ``Efficient document image
  classification using region-based graph neural network,'' 2021.

\end{thebibliography}

\end{document}